\documentclass[aps, amsfonts, amsmath, MnSymbol, nofootinbib, showpacs, byrevtex, 
preprintnumbers, showkeys, preprint, draft, prd]{article}   
\usepackage{amsmath}
\usepackage{epsf}
\usepackage{graphicx}

\newcommand{\cL}{{\cal L}}

\newcommand{\bi}{\bigskip}
\newcommand{\no}{\noindent}
\newcommand{\be}{\begin{equation}}
\newcommand{\ee}{\end{equation}}
\newcommand{\bea}{\begin{eqnarray}}
\newcommand{\eea}{\end{eqnarray}}

\newcommand{\rk}{\right)}
\newcommand{\lk}{\left(}

\newcommand{\sli}{\sum\limits}
\newcommand{\pli}{\prod\limits}
\newcommand{\il}{\int\limits}

\newcommand{\vE}{\vec{E}}

\newcommand{\vp}{\vec{p}}

\renewcommand{\vec}[1]{\mbox{\boldmath$#1$\unboldmath}}

\numberwithin{equation}{section}

\begin{document}

\title{On Schwinger's formula for pair production}

\author{W. Dittrich\\
Institut f\"ur Theoretische Physik\\
Universit\"at T\"ubingen\\
Auf der Morgenstelle 14\\
D-72076 T\"ubingen\\
Germany\\
electronic address: qed.dittrich@uni-tuebingen.de
}
\date{\today}
%


\maketitle
\bi

\no

\begin{abstract}
We present some comments on Schwingers's calculation of electron-positron production in a prescribed constant 
electric field. The range of validity of $2 Im {\cL}^{(1)} (E)$ is discussed thoroughly and 
limiting cases are provided.
\end{abstract}

\section{Number of electron-positron pairs produced in a uniform electric field}
\bigskip

\noindent
Start with Schwinger's expression for $2 Im {{\cL}^{(1)}} (E)$ \cite{schwinger}:
\bea
\label{1}
2 Im {{\cL}}^{(1)} = 2 \frac{(e E)^2}{(2 \pi)^3}  & &  \hspace{-1cm} \underbrace{\sli^\infty_{n = 1}   \frac{1}{n^2} e^{- \frac{\pi n}{\beta}}}_{} ,
 \hspace{0.5cm}\beta = \frac{eE}{m^2} \\ 
& = & Li_2 (e^{- \frac{\pi}{\beta}}) , \mbox{Euler's dilogarithm}  \nonumber
\eea
$\uparrow^z \vE = const.$\\
$\hbar = c = 1, \quad V = L^3$\\
In general for spin 
$s = \frac{1}{2}$ and $s = 0$:
\bea
2 Im {\cL}^{(1)} (E) &= &(2 s + 1) \frac{(e E)^2}{(2 \pi)^3} \sli^\infty_{n = 1} \frac{(\pm 1)^{n + 1}}{n^2} 
e^{- n \frac{\pi}{\beta}} \nonumber\\
&=& \pm (2 s + 1) \frac{(e E)^2}{(2 \pi)^3} Li_2 (\pm e^{- \frac{\pi}{\beta}}) \, .
\eea
Continuous phase space integration, how to count states:
\be
\int d^3 \vp \frac{V}{(2 \pi)^3} ./. = \il^{+ \infty}_{- \infty} \frac{L}{2 \pi} d p_1 
\il^{+ \infty}_{- \infty} \frac{L}{2 \pi} d p_2 \il^{+ \infty}_{- \infty} \frac{L}{2 \pi} d p_3 ./.
\ee
$t = \frac{p_3}{e E}$ in const. $E$-field: $d p_3 = e E dt, \quad \quad 0 \leq t \leq T$ or
$\il^{+ \infty}_{- \infty} d p_3 \to e E T$.
\bigskip

\noindent
Replace one factor $(e E)$ in (\ref{1}) by $\frac{1}{T} \il^{+ \infty}_{- \infty} d p_3$. Then
\be
\label{4}
Im {\cL}^{(1)} (E) T = \frac{1}{(2 \pi)^2} \il^{+ \infty}_{- \infty} d p_3 \frac{(e E)}{(2 \pi)} \sli^\infty_{n = 1}
\frac{1}{n^2} e^{- \frac{\pi n}{\beta}} \, .
\ee 
\bi

\no
Rewrite the sum in (\ref{4}):
\bea
\label{5x}
& & \frac{e E}{2 \pi} \sli^\infty_{n = 1} \frac{1}{n^2} e^{- \frac{\pi}{\beta} n} \nonumber\\
& = &  \sli^\infty_{n = 1}  
\frac{1}{n} e^{- \frac{\pi}{\beta} n} \frac{1}{2} \frac{1}{\pi \frac{n}{eE}} \, , \quad \quad 
\frac{1}{a} = \il^\infty_0 d x e^{- ax} , \quad \quad a = \pi \frac{n}{eE} \nonumber\\
& = &  \sli^\infty_{n = 1}  \frac{1}{n}  e^{- \frac{\pi}{\beta} n} \frac{1}{2} \il^\infty_0 d (p^2_\perp) e^{- \pi 
\frac{n}{eE} p^2_\perp} \, , \quad \quad d p^2_\perp = 2 p_\perp d p_\perp \, , \quad \quad \beta = \frac{eE}{m^2} 
\nonumber\\
& = &  \sli^\infty_{n = 1} \frac{1}{n} \il^\infty_0 p_\perp d p_\perp \exp \left\{ - \pi 
\frac{m^2 + p^2_\perp}{e E} n \right\} \nonumber\\
& = & \il^\infty_0 d p_\perp p_\perp \sli^\infty_{n = 1} \frac{1}{n}
\exp \left\{ - \pi \frac{m^2 + p^2_\perp}{e E} n \right\} \Longrightarrow \nonumber
\eea
Use $\ln (1 - x) = - \sli^\infty_{n = 1} \frac{1}{n} x^n \, , \quad \quad |x| < 1 \, , \quad \quad x = 
\exp \left\{ - \pi \frac{m^2 + p^2_\perp}{e E} \right\}$
\be
\label{5}
\Longrightarrow = - \il^\infty_0 d p_\perp p_\perp \ln (1   - e^{- \pi \lambda_p}) \Longrightarrow \, , \quad \quad \lambda_p = 
\frac{m^2 + p^2_\perp}{e E} 
\ee
\be
\label{6}
e^{- \pi \lambda_p} = e^{- \pi \frac{m^2 + p^2_\perp}{eE}} = \bar{n}_p \, .
\ee
This important expression - also probability for tunneling - relates the imaginary part of the Lagrangian of the 
field to the mean number $\bar{n}_p$ of electron-positron pairs produced by the field in the state with given momentum and spin
projection. $\bar{n}_p$ is degenerate  with respect to spin (two) and momentum $p_3$ with 
$\frac{L_3 \Delta p_3}{2 \pi \hbar}$ with $\Delta p_3 = e E T$.
\bi

\no
So we can continue to write (\ref{5}):
\be
\label{93}
\Longrightarrow - \il^\infty_0 d p_\perp p_\perp \ln (1 - \bar{n}_p) \, , \quad \quad \il^\infty_0 d p_\perp 2 \pi p_\perp ./.
= \il^\infty_{- \infty} d p_1 d p_2  ./. \nonumber
\ee
Here then is the relation between $Im {\cL}^{(1)} (E)$ and $\bar{n}_p$ (insert $\hbar$ and $V = L^3$):
\be
\label{7}
\frac{2}{\hbar} Im {\cL}^{(1)} V T = - 2 \int d^3 \vp \frac{V}{(2 \pi)^3} \ln (1 - \bar{n}_p)
\ee
\be
\label{8}
\bar{n}_p  = \exp \left\{ - \pi \frac{m^2 + p^2_\perp}{e E} \right\} \, .
\ee
This is Nikishov's virial representation of the imaginary part of ${\cL}^{(1)} (E)$ \cite{nikishov}.
\bi

\no
With the aid of (\ref{8}) let us prove Nikishov's result for the mean number of pairs in four-volume $VT$ by counting states:
\bea
\label{9}
\bar{n} & = & 2 \il^{+ \infty}_{- \infty} d p_1 \frac{L}{(2 \pi)} \il^{+ \infty}_{- \infty}  d p_2 
 \frac{L}{(2 \pi)} \il^{e E L}_{- \infty} d p_3 \frac{T}{(2 \pi)} \bar{n}_p \Longrightarrow \nonumber\\
& = & 2 \frac{L^2}{(2 \pi)^2} 2 \pi \il^\infty_0 p_\perp d p_\perp \frac{T}{2 \pi} e E L e^{- \pi \lambda_p } \nonumber\\
& = & \frac{(e E)}{(2 \pi)^2} V T \il^\infty_0 d p^2_\perp e^{- \pi \frac{p^2_\perp}{e E}} e^{- \frac{\pi}{\beta}} = 
\frac{e E}{(2 \pi)^2} V T \frac{1}{\pi \frac{1}{e E}} e^{- \frac{\pi}{\beta}} \, . \nonumber
\eea
Therefore
\be
\label{10}
\bar{n} = 2 \frac{(e E)^2}{(2 \pi)^3} V T e^{- \frac{\pi}{\beta}} \, , \quad \quad \beta = \frac{e E}{m^2} \, ,
\ee
which is Nikishov's result for the mean number of pairs produced in volume $V = L^3$ during time $T$.
For Bose particles the factor $2$ is suppressed.
\bi

\no
Introducing
\be
\label{11}
\xi = \exp \left\{ - \frac{\pi m^2}{e E} \right\} \, , \quad \quad \gamma = VT \frac{(e E)^2}{4 \pi^3} 
\ee
we also can write
\be
\label{12}
\bar{n} = \gamma \xi \, .
\ee
Formula (\ref{9}) is an approximation of the following expression for $n = 1$:
\bea
\label{13}
\int d^3 p \frac{V}{(2 \pi)^3} \sli^\infty_{n = 1} {\bar{n}^n_p}  \, .
\eea
Furthermore the well-known vacuum persistence probability $| \langle O_+ | O_- \rangle |^2$ is given by
\bea
\label{14}
| \langle O_+ | O_- \rangle |^2 & = & p_0 
 = \pli_{s, p} \lk {1} - e^{- \pi \lambda_p} \rk \nonumber\\
& = & \pli_{s, p} \lk 1 - e^{- \pi
\frac{m^2 + p^2_\perp}{e E}} \rk = e^{- 2 Im {\cL}^{(1)} (E) V T} \nonumber\\
& = & \exp \left\{ - VT \frac{(eE)^2}{4 \pi^3} \sli^\infty_{n = 1} \frac{1}{n^2} e^{- n \frac{\pi m^2}{eE}} \right\} \nonumber\\
& = & 
\exp \{ - \gamma Li_2 (\xi) \} \, .
\eea
Schwinger writes in his brilliant article \cite{schwinger} as well as in his 2nd volume on ``sources, particles
and fields'': ``We recognize in $2 Im {\cL}^{(1)}$ a measure in the {probability}, per unit time and unit spatial volume, 
that an electron-positron pair has been created.'' This statement is only true for very weak fields $(eE \ll m^2)$
in which case the contributions of the $n = 2, 3, \ldots$ terms in the sum of (\ref{1}) can be neglected:
\be
2 Im {\cL}^{(1)} (e E \ll m^2) = 2 \frac{(e E)^2}{(2 \pi)^3} e^{- \frac{\pi}{\beta}} \equiv \frac{\bar{n}}{VT} = \frac{\gamma \xi}{VT}
\simeq \frac{\gamma L}{VT} \, ; \quad L = - \ln (1 - \xi)  \, , \nonumber
\ee
which is identical to Nikishov's result. Here, in order to save at least part of Schwinger's statement, 
we are being a bit casual since $\bar{n}$ is not a probability but an average number.
 To be more specific let us start with (\ref{13}):
\bea
\label{166}
& &  2 \int d^3 \vp \frac{V}{(2 \pi)^3} \sli^\infty_{n = 1}  (\bar{n}_p)^n \, , 
\quad \quad \int d p_3 = T e E \nonumber\\
& = & V \frac{T eE}{4 \pi^3} \il^{+ \infty}_{- \infty} d p_1  \il^{+ \infty}_{- \infty} d p_2 \sli^\infty_{n = 1} 
(\bar{n}_p))^n \, , \quad \quad \int d p_1 \int dp_2 = 2 \pi \il^\infty_0 d p_\perp p_\perp \nonumber\\
& = & V T \frac{(e E)}{2 \pi^2} \il^\infty_0 p_\perp d p_\perp \sli^\infty_{n = 1}  e^{- \pi \lambda_p n} \, , \,
\bar{n}_p = e^{ - \pi \lambda_p} \, , \, \lambda_p = \frac{m^2 + p^2_\perp}{eE} \, , \,
\beta = \frac{e E}{m^2} \nonumber\\
& = & VT \frac{(e E)}{2 \pi^2} \sli^\infty_{n = 1}  e^{- \frac{\pi}{\beta} n} \frac{1}{2} \il^\infty_0 d p^2_\perp
e^{- \pi \frac{n}{eE} p^2_\perp} \nonumber\\
& =& V T \frac{(e E)}{2 \pi^2} \sli^\infty_{n = 1} e^{- \frac{\pi}{\beta} n} \frac{1}{2 \frac{\pi n}{e E}} = VT \frac{(e E)^2}{4 \pi^3} \sli^\infty_{n = 1} \frac{1}{n}
e^{- \frac{\pi m^2}{e E} n} \nonumber\\
& = & V T \frac{(e E)^2}{4 \pi^3} \left[ - \ln (1 - e^{- \pi \frac{m^2}{e E}}) \right] =: \frac{p_1}{p_0} = \gamma L \, . \nonumber
\eea
Finally
\bea
\label{15}
p_1 & = & V T \frac{(e E)^2}{4 \pi^3} \left[ - \ln \lk 1 - e^{- \pi \frac{m^2}{e E}} \rk \right] \cdot
\exp \left\{ - V T \frac{(e E)^2}{4 \pi^3} \sli^\infty_{n = 1} \frac{1}{n^2} e^{- n \frac{\pi m^2}{e E}} \right\} \nonumber\\
p_1 & = & \gamma L p_0 \, , \quad \quad L = - \ln (1 -\xi) \, , \quad \quad p_0 = \exp \left\{ - \gamma Li_2 (\xi) \right\} \, .
\eea
So far we have
\bea
\label{191}
p_0 & = & \exp \{ - \gamma Li_2 (\xi) \} \, , \quad \quad \mbox{Schwinger's vaccuum persistence probability} \nonumber\\
p_1 & =& \gamma L p_0 \, , \nonumber\\
\bar{n} & = & \gamma \xi \, . \nonumber
\eea
Let's denote by $\alpha = (\vp, s)$ the quantum numbers of the electron states. Then we can write our vacuum-to-vacuum probability as
\be
\label{16}
p_0 = \pli_\alpha (1 - \bar{n}_\alpha) \, .
\ee
In an electric field any number of pairs can be produced, so the probability that there are $n = 0, 1, 2, \ldots$ electron-positron 
pairs shows up in the series
\bea
\label{17}
\sli^\infty_{n = 0} p_n & = & \pli_\alpha (1 - \bar{n}_\alpha) + \sli_\alpha \bar{n}_\alpha \pli_{\beta \neq \alpha} (1 - \bar{n}_\beta)
\nonumber\\
& &  + 
\frac{1}{2!} \sli_{\alpha \neq \beta} \bar{n}_\alpha \bar{n}_\beta \pli_{\gamma \neq \alpha, \beta} (1 - \bar{n}_\gamma)
+ \ldots = 1 \, .
\eea
In our case, each of the quantities $p_n$ describes the propability
 for the number $n = 0, 1, 2, \ldots$ electron-positron pairs in four-volume. The first terms were calculated above:
\bea
\label{212}
p_0 & = & \exp \left\{ - \gamma Li_2 (\xi) \right\} \nonumber\\
p_1 & = & \gamma L p_0 \, .
\eea
The next and followers for the numbers $n$  were calculated by Krivoruchenko:
\be
\label{218}
p_2 = \frac{\gamma}{2} \lk \gamma L^2 + L - \frac{\xi}{1 - \xi} \rk p_0 \quad \quad \mbox{etc.}
\ee
It is highly interesting to follow Krivoruchenko's paper \cite{Krivoruchenko:2012rn} 
and find out that in electric fields of 
supercritical strength $| e E| > \frac{\pi m^2}{ln 2}$, the unitary condition (\ref{17}), \\
$\sli^\infty_{n = 0} p_n =~1$, changes into an asymptotic
divergence, i.a. the positive definiteness of  the probability is violated. This divergence indicates a failure of the continuum limit 
approximation, i.e. by the replacement of the discrete sum by the integral over the phase space:
\be
\label{226}
\sli_\alpha ./. \to 2 \int \frac{V}{(2 \pi)^3} d^3 \vp \ldots = VT |eE| \int \frac{2 d^2 p_\perp}{(2 \pi)^3} \ldots \, .
\ee
\bigskip

\bigskip
\bigskip

\bigskip

\noindent
\section{Schwinger's formula for $Im {\cL}^{(1)} (E)$ the long\\ way, i.e., without using the residue theorem.}
\bigskip

\noindent
Take the formula (5.27) or equivalently (6.33) of the ``Lecture Notes 220''
on ``Effective Lagrangians in QED'' by Dittrich and Reuter \cite{dittrich:1976,dittrich:1985}:
\begin{eqnarray}
{{\cL}}^{(1)} (B) & = & - \frac{1}{32 \pi^2} \left\{ (2 m^4 - 4 m^2 (eB) + \frac{4}{3} (e B)^2) \left[ 1 + \ln \lk \frac{m^2}{2 e B}
\rk \right] \right. \nonumber\\
& & \left. + 4 m^2 (e B) - 3 m^4 - (4 eB)^2  \zeta' \lk   - 1, \frac{m}{2 e B} \rk \right\} \, . 
\end{eqnarray}
This can also be written in the form
\begin{eqnarray}
& & {{\cL}}^{(1)} (B)\nonumber\\
 & = & - \frac{1}{32 \pi^2} \Big\{ - 3 m^4 + 4 (eB)^2 \lk \frac{1}{3} - 4 \zeta' (-1) \rk + 4 m^2 (eB) 
\ln 2 \pi - 1) 
\nonumber\\
&  & - 2 m^4 \ln \frac{2 e B}{m^2} - 4 m^2 (eB) \ln \frac{2 e B}{m^2} - \frac{4}{3} (eB)^2 \ln \frac{2 eB}{m^2} \nonumber\\
& &  - 16  (eB)^2 \il^{1+ \frac{m^2}{2 eB}}_1  d x \ln \Gamma (x) \Big\} \, .
\end{eqnarray}
Introducing the critical field strength $§B_{er} = \frac{m^2}{e}$ and measuring the magnetic field in this unit, we can rewrite the 
last expression as
\begin{eqnarray}
{{\cL}}^{(1)} (B) & =  & \frac{\alpha}{2 \pi} \Big\{ \frac{3}{4} - B (\ln (2 \pi) - 1) - B^2 \lk \frac{1}{3} - 4 
\zeta' (- 1) \rk  \nonumber\\
& &  + \lk \frac{1}{2} + B + \frac{1}{3} B^2 \rk \ln (2 B)   + 4 B^2 \il^{1 + \frac{1}{2 B}}_1 \ln  (\Gamma (x)) dx
\Big\} \, . 
\end{eqnarray}
For a pure electric field the Lagrangian is likewise given by
\begin{eqnarray}
{{\cL}}^{(1)} (E) & = & - \frac{1}{32 \pi^2} \left\{ \lk 2 m^4 + 4  i m^2 e E - \frac{4}{3} e^2 E^2 \rk 
\lk \ln \lk i \frac{m^2}{2 e E} \rk + 1 \rk \right. \nonumber\\
& & \left. - 3 m^4 - 4  i m^2 e E + 16 e^2 E^2 \zeta' \lk - 1, i \frac{m^2}{2 eE} \rk \right\} \, . 
\end{eqnarray}
It takes a litte practice to separate this formula into its real and imaginary part:
\begin{eqnarray}
{{\cL}}^{(1)} (E) & = & \frac{\alpha}{2 \pi} \Big\{ \frac{3}{4} + \frac{1}{2} \ln (2 E)  - \frac{\pi}{2} E + E^2 \lk
\frac{1}{3} - 4 \zeta' (- 1) \rk - \frac{1}{3} E^2 \ln (2 E)  \nonumber\\
& &  + 4  E^2 \il^{1/2 E}_0   Im \ln (\Gamma (1 + y)) dy \Big\} \nonumber\\
& & + i \frac{\alpha}{2 \pi} \Big\{ - \frac{\pi}{4} - E \ln (2 E) + E (\ln (2 \pi) - 1) \nonumber\\
& & + \frac{\pi}{6} E^2 - 4 E^2 
\il^{1/2 E}_0 Re \ln (\Gamma (1 + i y)) d y \Big\} \, . 
\end{eqnarray}
With the aid of the relation (Gradshteyn / Ryzhik)
\begin{eqnarray}
Re \ln (\Gamma (1 + i y)) & = & \ln | \Gamma (1 + iy)| = \frac{1}{2} \ln \left[ \Gamma (1 + i y) \Gamma (1 - iy) \right]\nonumber\\
& = & - \frac{1}{2} \ln \frac{\sinh (\pi y)}{\pi y} \nonumber
\end{eqnarray}
and an integration by parts
\begin{eqnarray}
\il^{1/2 E}_0 1 \cdot \ln \frac{\sinh (\pi y)}{\pi y} d y = \frac{1}{2 E} \ln \left[ \frac{2 E}{\pi} \sinh \frac{\pi}{2 E}
\right] - \il^{1/2 E}_{0} \left[ \pi y \coth (\pi y) -1 \right] d y \, , \nonumber
\end{eqnarray}
we obtain
\begin{eqnarray}
\label{443-2.6}
 Im {{\cL}}^{(1)} (E) 
& = & \frac{\alpha}{2 \pi} \Big\{ - \frac{\pi}{4} + E \ln 2 + \frac{\pi}{6} E^2 + E \ln 
\lk \sinh \frac{\pi}{2 E} \rk\nonumber\\
&& - 2 E^2 \il^{1/2 E}_{0} \pi y \coth (\pi y) dy \Big\} \, .
\end{eqnarray}
Let's change the variable $x = \pi y$ and evaluate the integral on the right-hand side with the use of [Wolfram Mathematica online 
integrator]
\begin{eqnarray}
\int x \coth (x) d x & =& \frac{1}{2} \lk x (x + 2 \ln (1 - e^{- 2x})) - Li_2 (e^{- 2x}) \rk \nonumber\\
& = & \frac{1}{2} \lk x (x + 2 \ln (2 e^{- x} \sinh (x)) - Li_2 (e^{- 2x}) \rk \, . 
\end{eqnarray}
With integration limits we arrive at
\begin{eqnarray}
 \frac{1}{\pi} \il^{\frac{\pi}{2} E}_0 x \coth (x) d x 
 & = & \frac{1}{2 \pi} \Big[ \frac{\pi}{2 E} 
\lk \frac{\pi}{2 E} + 2 \ln 2 
- \frac{\pi}{E} + 2 \ln \lk \sinh \lk \frac{\pi}{2 E} \rk \rk \rk \nonumber\\
& & - Li_2 \lk e^{- \frac{\pi}{E}} \rk + \frac{\pi^2}{6} 
\Big] \, . 
\end{eqnarray}
We substitute this in our last expression for $Im {{\cL}}^{(1)} (E)$ and obtain
\begin{eqnarray}
Im {{\cL}}^{(1)} (E) & =& \frac{\alpha}{2 \pi} \left\{ - \frac{\pi}{4} + E \ln 2 + \frac{\pi}{6} E^2 + E \ln 
\lk \sinh \lk \frac{\pi}{2 E} \rk \rk \right. \nonumber\\
& & \left. + \frac{\pi}{4} - E \ln 2 - E \ln \lk \sinh \lk \frac{\pi}{2 E} \rk \rk + \frac{E^2}{\pi} Li_2 \lk e^{- \frac{\pi}{E}} \rk - \frac{\pi}{6} E^2
\right\} \nonumber\\
& =  & \frac{\alpha}{2 \pi} \frac{E^2}{\pi} Li_2 \lk e^{- \frac{\pi}{E}} \rk \quad \quad \nonumber\\
\Big( Li_2 \lk e^{- \frac{\pi}{E}} \rk & = & \sli^\infty_{n = 1} \frac{1}{n^2} e^{- \frac{\pi}{E} n} \, ; \nonumber\\
& & 
 \mbox{in 
units of} \quad E_{cr} = \frac{m^2}{e} \quad : \quad \sli^\infty_{n = 1} 
\frac{1}{n^2} e^{- \frac{m^2 \pi}{e E} n} \Big) \, .
\end{eqnarray}
Finally we obtain J.S.'s famous result $(\alpha = \frac{e^2}{4 \pi})$:
\begin{eqnarray}
Im {{\cL}}^{(1)} (E) = \frac{\alpha}{2 \pi^2} E^2 \sli^\infty_{n = 1} \frac{1}{n^2} e^{- \frac{m^2 \pi}{e E} n} \, .
\end{eqnarray}
At last we might add the result for the real part  \cite{Soldati:2011gi}:
\begin{eqnarray}
Re \cL^{(1)} (E) & = & - \frac{e^2 E^2}{4 \pi^4} \left( C + \ln \frac{\pi m^2}{eE} \right) \sli^\infty_{n = 1} \frac{1}{n^2} \cosh \left\{
n \frac{\pi m^2}{e E} \right\} \nonumber\\
& & - \frac{e^2 E^2}{4 \pi^4} \sli^\infty_{n = 1} \frac{\ln n}{n^2} \cosh \left\{ n \frac{\pi m^2}{e E} \right\}  \, ,
\end{eqnarray}
where $C$ is the Euler-Mascheroni constant.

\section{Acknowledgement}
The author wishes to thank M. I. Krivoruchenko for critical remarks and several interesting discussions.

\end{document}